# Compact polarization-independent non-volatile optical switches

FANGFA BAO,[1] JIAKAI RUAN,[1] WEIJIA LI,[2] WEI ZHANG,[1] GUOXIANG WANG,[1] XIANG SHEN[1] AND YIXIAO GAO[1,*]

[1]*Laboratory of Infrared Materials and Devices, Zhejiang Key Laboratory of Advanced Optical Functional Materials and Devices, Research Institute of Advanced Technologies, Ningbo University, Ningbo 315211, China*
[2]*Dept. of Electrical and Computer Engineering, McGill University, Montreal, QC H3A 0G4, Canada*
*gaoyixiao@nbu.edu.cn*

**Compact, non-volatile optical switches on silicon platforms are essential for reconfigurable photonics, but the strong anisotropy of silicon waveguides leads to polarization-dependent performance. In this paper, we propose a polarization-independent, non-volatile optical switch utilizing low-loss phase change material (PCM) $Sb_2S_3$. By incorporating $Sb_2S_3$ into a multimode slot waveguide, multimode interference can be efficiently tuned for both TE and TM polarizations, owing to enhanced light–PCM interaction. Polarization-independent switching is achieved through the optimal design of the multimode slot waveguide region. The proposed non-volatile switch demonstrates a crosstalk (CT) < -21.9 dB and insertion loss (IL) < 0.12 dB at 1550 nm with a multimode section length of 9.67 μm, which may find promising applications in reconfigurable photonic circuits for on-chip optical signal processing.**

Reconfigurable integrated photonic circuits with energy-efficiency and compact footprint are highly desired for on-chip optical communication and photonic computing technologies [1,2]. Silicon photonics becomes the major platform for on-chip devices with the advantages of high-density integration and low-cost manufacturing. However, traditional silicon photonic devices rely on carrier or thermo-optic (TO) effect to control refractive index, which have a limited tuning range (e.g. TO coefficient of silicon is $1.8\times10^{-4}$ /K [3]), and thus reconfigurable silicon devices usually requires a long device length on the order of hundreds of microns. Additionally, maintaining the refractive index changes requires continuous energy supply, leading to huge power consumption [4].

Phase-change materials could stably maintain in their crystalline (Cr) or amorphous (Am) state at room temperature, and non-volatile phase transitions can be induced optically or electrically, accompanied by significant changes in refractive index [5]. Benefiting from back-end CMOS compatibility, PCMs could hybrid-integrated onto silicon waveguides to actively tune the modal properties [6], making them a promising functional material for reconfigurable silicon photonics [7–9]. However, traditional PCMs (e.g. $Ge_2Sb_2Te_5$ and $Ge_2Sb_2Se_4Te_2$ [10]) have large extinction coefficients in their Cr state, leading to high insertion losses. Recently, binary-element PCMs including $Sb_2S_3$ and $Sb_2Se_3$ offer refractive index contrasts of $\Delta n \sim 0.6$ upon phase change and extremely low extinction coefficients [11], making them ideal material choices for purely phase tuning capabilities, which are widely explored to construct compact, low-insertion-loss, energy-efficient reconfigurable photonic devices, including optical switches [8,12,13], tunable filters [14,15], mode-division multiplexer [16], polarization-handling devices [17,18], etc.

As key components of reconfigurable photonic circuits, silicon switches are usually polarization sensitive. Owing to the rectangular cross-section of waveguides, the transverse electric (TE) and transverse magnetic (TM) modes have distinct effective refractive indices ($n_{eff}$), leading to different transmission behavior between two polarizations. The demand for polarization-independent optical switches is important for practical applications [19]. Wang et al. utilized a low-index-contrast waveguide to reduce the $n_{eff}$ difference between TE and TM modes so as to realize a polarization-independent switching, while such scheme cannot be employed in the high-index contrast silicon photonics [20]. By cascading a polarization converter with single-polarization switching structures [21–23], polarization-independent switching can be achieved, although this approach results in a larger device footprint. Wang et al. proposed a thermally-tuned, polarization-independent switch based on a Mach-Zehnder structure containing a polarization-insensitive multimode interference beam splitter [24], indicating a fine-engineered multimode waveguide could achieve a polarization-independent operation. However, a non-volatile optical switch with polarization-independent feature and compact footprint remains unexplored.

In this paper, we propose a compact polarization-independent non-volatile optical switch enabled by ultralow loss PCM $Sb_2S_3$. Despite a moderate refractive contrast ($\Delta n \sim 0.6$) of $Sb_2S_3$, efficient modal field manipulation is achieved by incorporating $Sb_2S_3$ into a slot waveguide. By controlling phase-change-mediated multimode interference within the multimode slot waveguide, a polarization-independent switch could be achieved with low crosstalk (CT) and insertion loss (IL), making it a promising candidate for applications in reconfigurable photonics.

Figure 1 shows the schematic of the proposed polarization-independent 2 × 2 optical switches. The switch consists of a multimode slot waveguide with low-loss $Sb_2S_3$ sandwiched in the slot region, and four single-mode waveguides as input/output ports connecting to the multimode waveguide. The basic working principle is illustrated in Figure 1(a): the injected mode from one input port will output to cross port with crystalline $Sb_2S_3$ (c-$Sb_2S_3$) or to bar port with amorphous $Sb_2S_3$ (a-$Sb_2S_3$), where TE or TM mode input have the same transmission feature with the help of an optimally designed multimode region which will be discussed later. The whole device is designed based on a silicon on insulator platform with a 300 nm thick silicon top layer ($h$ = 300 nm). The length and width of the multimode waveguide are $L$ and $W_m$, and the slot width is $W_{ss}$, as depicted in Fig. 1(b). The width of single mode waveguide is half of $W_m$. The refractive index of $Sb_2S_3$ is obtained from Ref. [11]. The input and output single-mode waveguides are designed to have an S-bend shape as shown in Figure 1(c), with a spacing of $d$ between the two input (output) waveguides and considering the total length of the S-bend as $L_s$. Here we consider $L_s$ = 8 μm and $d$ = 4 μm respectively. Note that microheaters based on graphene [25], PIN diodes [26], and ITO [27] can effectively trigger phase transitions in integrated photonic circuits. In the proposed design, the phase change in $Sb_2S_3$ can be similarly activated using external microheaters (see Supplement 1).

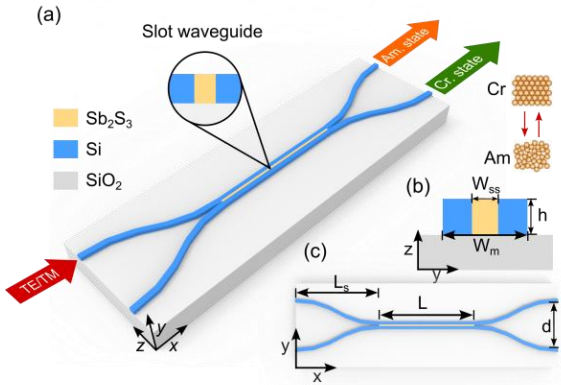

**Figure 1** (a) Schematic of the proposed polarization-independent non-volatile optical switch (b) Cross-section of the multimode slot waveguide. (c) Top view of the optical switch.

Manipulation of the multimode interference is essential for the proposed optical switch. We begin by presenting the mode properties of the $Sb_2S_3$-integrated multimode slot waveguide. Figure 2(a) shows the mode profiles of TE and TM modes. The fundamental modes, i.e. $TE_0$ and $TM_0$ modes, are symmetric with respect to the waveguide center, while the second order modes including $TE_1$ and $TM_1$ modes are anti-symmetric, manifested by a field node in the center. When the multimode region only supports two lowest orders of modes for both polarizations, the co-propagating of the symmetric and antisymmetric modes with the same polarization and equal amplitudes could lead to a beat phenomenon owing to a multimode interference, as depicted in Fig. 2(b). The period of the beat pattern, or the coupling length $L_c$, is determined by the difference of propagation constants of two modes, as $L_c = \lambda/2(n_{eff0} - n_{eff1})$, where $n_{eff0}$ and $n_{eff1}$ is the effective mode index of fundamental and second order modes, and $\lambda$ is the working wavelength.

When we construct a polarization-independent optical switch, the coupling length $L_c$ and the length of the multimode region $L$ have to satisfy the following relation:

$$L = pL_{c,TE}^{Am} = (p+m)L_{c,TM}^{Am} = qL_{c,TM}^{Cr} = (q+n)L_{c,TE}^{Cr} \quad (1)$$

where $m$ and $n$ are even numbers but $p$ has the opposite parity to $q$. The physical picture of the parity-based condition in Eq. (1) is that, when light is injected into the multimode region, for one certain phase state, TE and TM modes would experience the same parity number (e.g. an odd number) of beats and then output from the same output port. When phase change occurs, TE and TM modes coupled to another output port after a opposite parity number (e.g. an even number) of beats. The challenge lies in simultaneously controlling multimode interference for both polarizations, involving the interaction of four modes in this process.

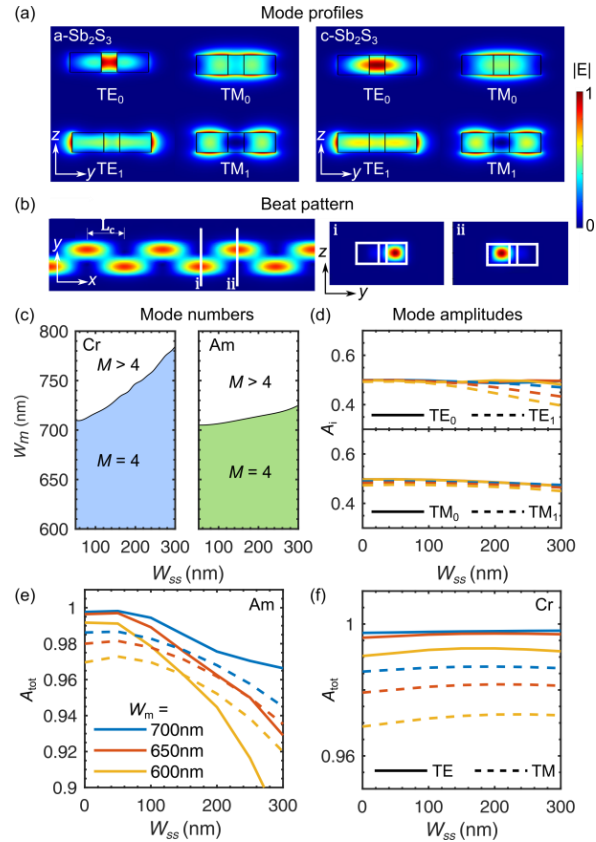

**Figure 2.** (a) TE and TM mode profiles in the slot waveguide at 1550 nm. (b) Multimode interference within the slot waveguide. $L_c$ denotes the coupling length. The cross-sectional field profiles at position i and ii is shown on the right. (c) Mode number $M$ mapping in the $W_m$-$W_{ss}$ parametric space. The blue and green regions correspond to the two-mode region of both polarizations. (d) Modal excitation coefficient of $TE_0$/$TE_1$ and $TM_0$/$TM_1$ modes at the junction of single-mode waveguide and multimode slot waveguide as a function of $W_{ss}$. The summation of mode excitation coefficient $A_i$ of TE and TM modes as a function of $W_{ss}$, when $Sb_2S_3$ is in (e) Am and. (f) Cr state.

Ensuring the multimode region only supports the two lowest-order modes for either polarization is key to minimize CT. The number of modes $M$ in the slot waveguide depends on $W_m$, $W_{ss}$, as well as the phase of $Sb_2S_3$. Figure 2(c) shows the mode number map supported in the slot waveguide in $W_m$-$W_{ss}$ space. The blue and green regions correspond to the two-mode region with Am and Cr phase, respectively. We focus on the overlapping region, where we restrict $W_m$ to the range of 600 nm to 700 nm and $W_{ss}$ to the range of 50 nm to 300 nm for the subsequent discussion.

Next, we design the cross-section of the slot waveguide, considering two key factors: 1) the TE (TM) input should be efficiently and equally coupled into two TE (TM) modes with opposite symmetry in the multimode slot waveguide to minimize IL and CT; 2) The TE and TM input modes should be directed to the same output port and switched to the other port upon phase change of $Sb_2S_3$, enabling polarization-independent switching.

The first factor can be addressed using mode expansion technique, i.e., any stable electromagnetic field **E** in a waveguide can be decomposed into a set of orthogonal waveguide modes, as $\mathbf{E} = \sum_i a_i \mathbf{e}_i \exp(i\beta_i x)$ where $a_i$, $\mathbf{e}_i$, and $\beta_i$ are the modal amplitude, normalized electric field and propagation constant of $i$-th mode, respectively. Figure 2(d) shows the relationship between the modal excitation coefficient $A_i$ of $TE_0/TE_1$ and $TM_0/TM_1$ modes in a slotted waveguide and $W_{ss}$ with amorphous $Sb_2S_3$, where $A_i = |a_i|^2/|a_0|^2$ with $a_0$ is the modal amplitude in the single mode waveguide, and $A_{tot} = \Sigma A_i$ with $i$ = 1, and 2 denoting fundamental and second order modes. Within the range of 50-150 nm, the fundamental and second-order modes can be effectively excited with approximately equal amplitudes. A similar result could also be obtained when $Sb_2S_3$ crystallized. Figure 2(e-f) shows the modal excitation coefficient $A_i$ at various $W_m$ and $W_{ss}$ values when $Sb_2S_3$ is in its two phases states. We could observe the tendency that, with a wider slot waveguide width, i.e., a larger $W_m$, $A_{tot}$ would be larger, indicating the input light would be predominately coupled into the two desired modes in the slot waveguide. However, for Am state, $A_{tot}$ would drop significantly with a larger slot width, indicating a stronger scattering loss at the junction between single mode and slot waveguide. This is attributed to the larger refractive index contrast between silicon and a-$Sb_2S_3$, compared with c-$Sb_2S_3$. Therefore, we choose $W_m$ equals to 700 nm and $W_{ss}$ should be less than 150 nm, in order to improve the coupling efficiency into desired modes and reduce the scattering loss of TM modes.

The second factor could be addressed by adjusting slot width $W_{ss}$ to satisfy Eq. (1). It should be noted that the four input and output single mode waveguide would couple with each other near the multimode region due to small spacing between two input (output) waveguides, therefore we cannot simply use Eq. (1) to determine the length of the multimode region. To address this issue, we define a length of the multimode region when the input mode couple into multimode region and beat twice and then coupled to the bar port, as the transition length $L_{tr}$, as depicted in Fig. 3(a). Thus, we need to introduce $L_{tr}$ into Eq. (1) to account for the additional coupling outside the multimode region. For example, when we consider TE mode and amorphous $Sb_2S_3$ phase, $L = L_{tr,TE}^{Am} + pL_{c,TE}^{Am}$. Note that $L_{tr}$ is also dependent on the polarization, phase state as well as the waveguide cross-section parameters. In order to simplify the calculation, we employed an approximate yet effective method to estimate $L_{tr}$. Figure 3(b) show the CT between two output ports as a function of multimode region length $L$, where two modes only beat once. $L_{tr}$ can be determined by evaluating the minimum CT. Here $W_m$=700nm and $W_{ss}$ equals 50 nm and 150 nm, and $L_{tr}$ at one certain polarization and phase state is obtained by averaging $L_{tr}$ at 50 nm and 150 nm, and thus we have $L_{tr,TE}^{Am}$=1.025 μm, $L_{tr,TM}^{Am}$=1.15 μm,

$L_{tr,TE}^{Cr}$ =0.975 μm, $L_{tr,TM}^{Cr}$ =1.025 μm. Figure 3(c) shows the multimode slot waveguide length as a function of beat times $X$ for two polarization state when $Sb_2S_3$ in its Am and Cr states, when $W_m$ = 700nm and $W_{ss}$=140 nm. We find $L$ = 9.67 μm produces the best approximation result satisfying Eq. (1), where $p$ = 4, $q$ = 5, $m$ = 2, and $n$ = 2. This corresponds to 4 couplings for TE mode, 6 couplings for TM polarization when $Sb_2S_3$ is in amorphous, and 5 couplings for TE polarization, and 7 couplings for TM polarization when $Sb_2S_3$ crystallized.

To summarize the design process, achieving optimal polarization-independent switching performance via multimode interference control involves three key steps: (1) Ensuring equal excitation amplitudes of the fundamental and second-order modes for both polarizations by engineering the junction between the input waveguide and the multimode slot waveguide; (2) Matching the coupling lengths of TE and TM modes by carefully tuning the cross-sectional dimensions of the multimode slot waveguide in order to satisfy the condition in Eq. (1); (3) Compensating for additional coupling effects introduced by the S-bend input waveguides through the incorporation of a transition length $L_{tr}$.

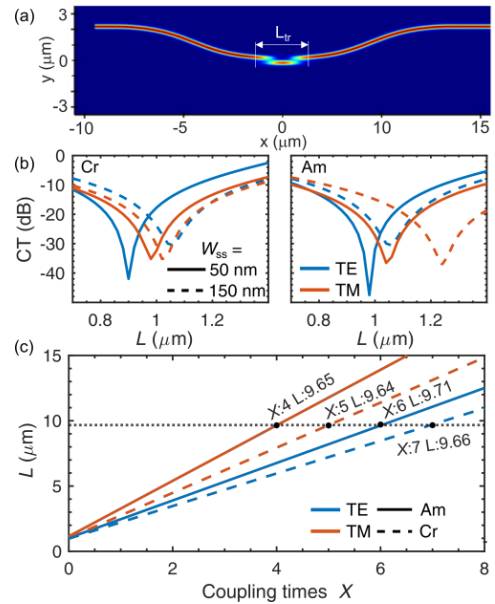

**Figure 3** (a) Definition of the transition length $L_{tr}$. (b) The CT of output ports as a function of the length of multimode slot waveguide at the wavelength of 1550 nm. Solid and dashed curves correspond to $W_{ss}$ = 50 nm and 150 nm, respectively. TE and TM mode are indicated by blue and red colors. (c) The change in the physical length of the multimode region, $L$, with respect to the coupling times $X$ for the two polarizations with amorphous (solid) and crystalline $Sb_2S_3$(dashed). The dotted line indicates the optimal $L$ equaling to 9.67 μm.

We validated our design through 3D FDTD simulation. Here the parameters of the switch are set as $L$ = 9.67 μm, $L_s$ = 8 μm, $d$ = 4 μm, $W_m$ = 700nm, $W_{ss}$ = 140nm, $h$ = 300nm. Figure 4(a) shows the electric field distribution. At 1550-nm-wavelength mode is injected into the single mode waveguide from the upper left port. When $Sb_2S_3$ is amorphous, both TE and TM polarized modes are output from the bar port, with TE mode coupling 8 times and TM mode coupling 6 times. When $Sb_2S_3$ is phase changed into crystalline state, both TE and TM modes are output from cross port, indicating the designed optical switch indeed exhibit a polarization-independent performance. Figure 4(b) shows transmission spectra at bar and cross output port within the wavelength range from 1530 nm to 1565 nm. At the wavelength of 1550nm, the CT of TE/TM modes in Am and Cr states is -25.3 dB/-26.8 dB and -21.9 dB/-27.4 dB, respectively, with IL of 0.09 dB/0.02 dB and 0.04 dB/0.08 dB. The designed proposed optical switch also demonstrates a capability of broadband operation, with an overall CT less than -9.3 dB, and an overall IL of 0.56 dB. Note that the presence of an external microheater may alter the mode properties; however, by applying the same design methodology, optimized polarization-independent switching performance can still be achieved (see Supplement 1).

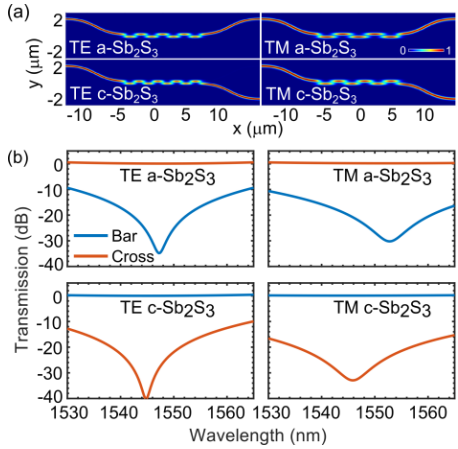

**Figure 4.** (a) Normalized electric field distribution at 1550nm. (b) transmission spectra of the proposed switches with TE and TM inputs.

Table 1 presents a performance comparison between our design and previously reported polarization independent switches. The proposed design has a device length of only 25.67 μm, which is the smallest compared to previous works. Furthermore, our design achieves low IL, benefiting from the ultralow-loss properties of $Sb_2S_3$. It is worth noting that recent advances in nanofabrication have made it feasible to integrate PCMs into narrow slot regions. For example, Refs. [28,29] report the successful integration of 75-nm-thick GST into a 70 nm-wide metallic gap, while Ref. [30] demonstrates the deposition of $VO_2$ into the gap of a silicon waveguide. These results indicate that the fabrication of the proposed switch is technically achievable. We also performed a fabrication tolerance analysis in Supplement 1.

**Table 1. Performance comparison of optical switches**

| Study type | Platform | Tuning method | IL (dB) | CT(dB) | Device Length |
|---|---|---|---|---|---|
| Exp. [20] | Polymer | Thermal | 0.6 | -30 | 22.5 mm |
| Sim. [22] | SOI | GST | 1.13 | -20 | ~160 μm |
| Exp. [23] | LNOI | Electrical | 4 | -10 | ~3 mm |
| Exp. [21] | SOI | Thermal | 2.3 | -15 | 1090 μm |
| Exp. [24] | SOI | Thermal | 1.8 | -30 | ~500 μm |
| Exp. [31] | SOI | Thermal | 0.6 | -30 | ~160 μm |
| Our work | SOI | Sb2S3 | 0.12 | -21.9 | 25.67 μm |

In conclusion, we propose a non-volatile, polarization-independent optical switch based on $Sb_2S_3$-mediated multimode interference. By precisely tailoring the effective refractive indices of the TE and TM modes, the coupling lengths of both polarizations are engineered to satisfy the condition defined in Eq. (1), enabling them to share the same output port and allowing simultaneous switching via a phase change. Owing to the strong interaction between the optical modes and PCM, polarization-independent switching is realized using a multimode slot waveguide with a compact length of 25.67 μm. At a wavelength of 1550 nm, the CT for TE/TM modes is −25.3 dB/−26.8 dB in the Am state and −21.9 dB/−27.4 dB in the Cr state, with corresponding IL of 0.09 dB/0.02 dB and 0.04 dB/0.08 dB. Furthermore, the device demonstrates broadband operation, maintaining CT below −9.3 dB and IL under 0.56 dB across the telecom C band, which shows great promise for on-chip optical communication and computing applications.

**Funding.** National Natural Science Foundation of China (62105172), the Youth Science and Technology Innovation Leading Talent Project of Ningbo City, China (2023QL005)

**Disclosures.** The authors declare no conflicts of interest.

**Data availability.** The data of this study are available from the corresponding authors upon reasonable request.

**Supplemental document**. See Supplement 1 for supporting content.

# COMPACT POLARIZATION-INDEPENDENT NON-VOLATILE OPTICAL SWITCHES: SUPPLEMENTAL DOCUMENT

## 1. Thermal simulation of the ITO microheater for triggering phase transitions

Phase-change behavior is typically governed by thermal effects. In addition to optical excitation methods [1], external electrical microheaters provide a flexible and widely adopted approach for controlling the phase state. These microheaters are commonly implemented using PIN structures [2], graphene [3], or ITO thin films [4], where electrical current generates Joule heating to induce the phase transition. In this work, we adopt an ITO-based microheater as a representative example due to its compatibility with CMOS fabrication processes. To assess its thermal performance, we performed finite element simulations using COMSOL Multiphysics, following the modeling framework reported in Ref. [5-6].

Figure S1 illustrates the configuration of the ITO microheater. From bottom to top, the structure consists of a silica substrate, a silicon waveguide, a phase-change layer of $Sb_2S_3$, an ITO heating layer, palladium metal electrodes, and a protective silica cladding on the top surface (not shown in the figure). The palladium electrodes, selected for their excellent conductivity and thermal stability, are 50 nm thick and placed on both sides of the ITO heater to deliver electrical pulses for Joule heating. Material parameters used in our simulation is summarized in Table S1.

$Sb_2S_3$ exhibits a crystallization temperature $T_c$ = 573 K and a melting temperature $T_m$ = 823 K [7]. The material parameters used in the thermal simulation are summarized in Table 1. Crystallization occurs when the PCM is heated above $T_c$ and maintained for a sufficient duration (e.g. tens of microseconds [8]). In contrast, amorphization is more technically demanding, as it requires heating above $T_m$ followed by rapid quenching to room temperature at a rate typically exceeding 1 K/ns [9], to preserve the disordered atomic arrangement.

As shown in Figure S2, our simulation results demonstrate that applying an 8 V voltage pulse with a duration of 25 ns raises the PCM temperature above $T_m$. The temperature then drops from $T_m$ to $T_c$ within approximately 135 ns, giving a quenching time $t_q$=135 ns. The corresponding quenching rate $R_q = (T_m - T_c)/t_q$ exceeds the typical 1 K/ns threshold throughout the cooling process, confirming that the ITO microheater is capable of achieving both crystallization and amorphization of $Sb_2S_3$.

Therefore, the proposed polarization-independent optical switch can be reliably reconfigured using an integrated ITO microheater, demonstrating its practical feasibility for non-volatile photonic applications.

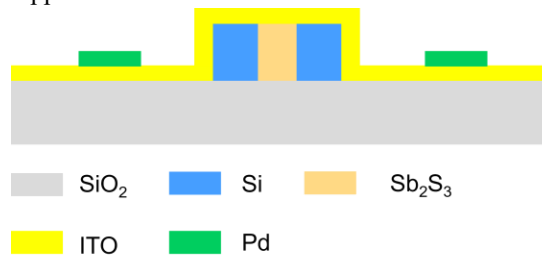

**Figure S1** Schematic of the thermal simulation model

Table S1. Material parameters for thermal simulation

| Material | Density (kg m$^{-3}$) | Heat capacity (J/kg K) | Thermal conductivity (W/mK) | Conductivity (S/m) |
|---|---|---|---|---|
| Sb$_2$S$_3$ | 4600 | 353 | 1.2 | / |
| ITO | 7100 | 1340 | 3.2 | 62500 [10] |
| Pd | 12023 | 244 | 71.8 | σ(T) from COMSOL |
| Si | 2329 | 713 | 140 | / |
| SiO$_2$ | 2203 | 746 | 1.38 | / |

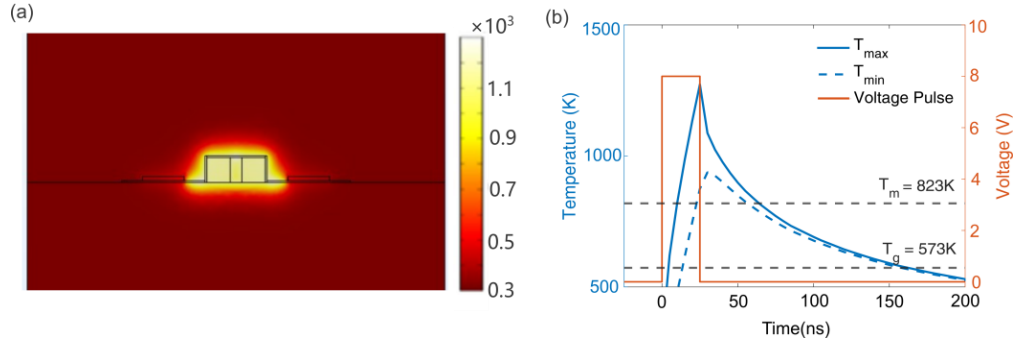

**Figure S2** (a) Simulated temperature distribution in the slot waveguide at the end of the electrical pulse used for amorphization. (b) FEM-simulated transient temperature response under an applied voltage pulse ($U$ = 8 V, $t$ = 25 ns) for inducing amorphization.

## 2. Design of the polarization-independent optical switch with an ITO microheater

The use of an external ITO microheater to trigger the phase change may alter the mode properties. Here, we demonstrate that optimized polarization independent switching performance can still be achieved by applying the same design methodology described in the main text.

Here, the ITO layer was modeled with a complex refractive index of 1.9615 + 0.00559i and a thickness of 20 nm [11], conformally covering the multimode region. Following the design procedure in the main text, the optimized device parameters with the ITO microheater are: $L$ = 9.5 μm, $W_m$ = 685 nm, and $W_{ss}$ = 140 nm. Figure S3(a) presents the normalized electric field distribution at a wavelength of 1550 nm, with the mode injected into the single-mode waveguide from the upper-left port. The simulated transmission spectra are shown in Fig. S3(b). When Sb$_2$S$_3$ is in the amorphous state, the CT and IL of the TE/TM modes at 1550 nm are −32.0 dB/−28.5 dB and 0.23 dB/0.36 dB, respectively. When Sb$_2$S$_3$ is crystallized, the CT and IL values for TE/TM input are −29.9 dB/−26.8 dB and 0.11 dB/0.23 dB, respectively. Within the operating bandwidth of 1530–1565 nm, the overall CT and IL remain below −8 dB (−11.2 dB) and 0.85 dB (0.67 dB) for the TE (TM) modes.

Note that the IL increases slightly due to the intrinsic absorption of ITO near 1550 nm. Nevertheless, owing to the compact multimode region (length of 9.5 μm) enabled by the high refractive index contrast of $Sb_2S_3$ and the strong PCM–mode interaction in the slot waveguide, the additional IL induced by ITO layer remains within an acceptable range.

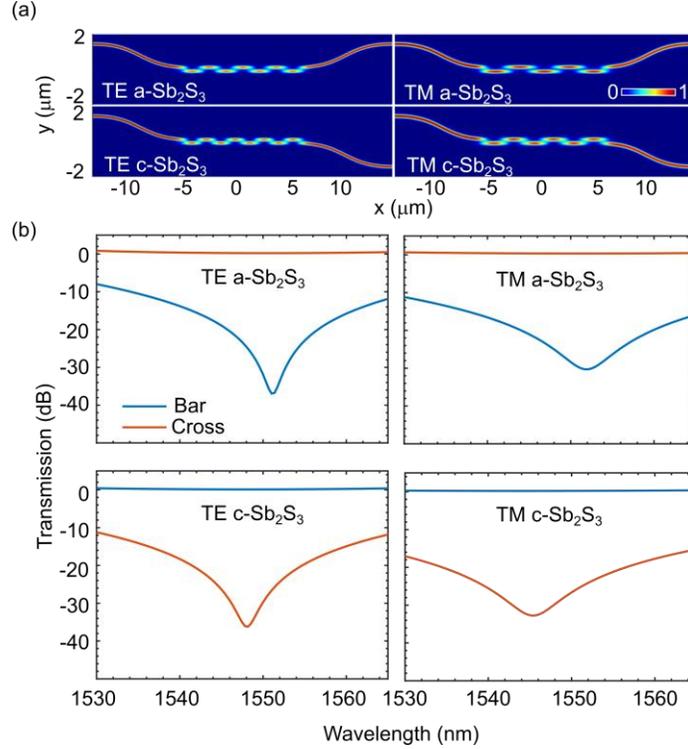

**Figure S3** (a) Normalized electric field distribution at 1550 nm with light injected into the single-mode waveguide from the upper-left port. (b) Transmission spectra of the proposed optical switch with an ITO microheater when $Sb_2S_3$ is in the amorphous and crystalline states.

### 3. Device tolerance

We have conducted a comprehensive analysis of fabrication tolerances to evaluate their potential impact on the device performance. Specifically, we consider three types of typical fabrication deviations: (1) the thickness variation of the $Sb_2S_3$ layer arising from the PCM deposition process, (2) slot width deviation and (3) waveguide width variation resulting from pattern transfer and etching inaccuracies.

**(1) The influence of $Sb_2S_3$ thickness variation**

During film deposition, the $Sb_2S_3$ layer may deviate from its designed thickness. To assess the effect of this variation, we simulated the insertion loss (IL) and crosstalk (CT) while varying the $Sb_2S_3$ thickness by $\Delta h_s$ within ±10 nm, as indicated by shaded region in Fig. S4(a). The results show that CT remains below -16.3 dB and IL stays under 0.2 dB across the $\Delta h_s$ range, suggesting the device exhibits good robustness to $Sb_2S_3$ thickness fluctuations.

**(2) The influence of slot width variation**

Slot width deviations ($\Delta w_{ss}$) can occur during the etching process. We evaluated the switching performance with the slot fully filled with $Sb_2S_3$ while varying $\Delta w_{ss}$ within ±5 nm, as shown in Fig. S4(b). The results indicate that CT remains below −20.3 dB and IL is under 0.18 dB throughout the examined range, indicating that the device is relatively insensitive to small variations in slot width.

**(3) The influence of waveguide width variation**

Waveguide width deviations ($\Delta w$), often resulting from lithography or etching errors, were also analyzed. In this case, we fixed the center-to-center spacing of the waveguide pair (420 nm) and varied the individual waveguide width $\Delta w$, as illustrated in Fig. 4(c). The simulation assumes the slot remains fully filled with $Sb_2S_3$. The results show that both CT and IL are sensitive to $\Delta w$: CT degrades significantly, and IL can increase to ~2 dB when $|\Delta w|$ approaches 5 nm. This indicates that precise control over the waveguide width is critical for maintaining optimal device performance.

In summary, our tolerance analysis reveals that the performance of the proposed optical switch is relatively robust to small variations in $Sb_2S_3$ thickness and slot width. However, waveguide width deviations have a significant impact, underscoring the need for high-precision control during fabrication.

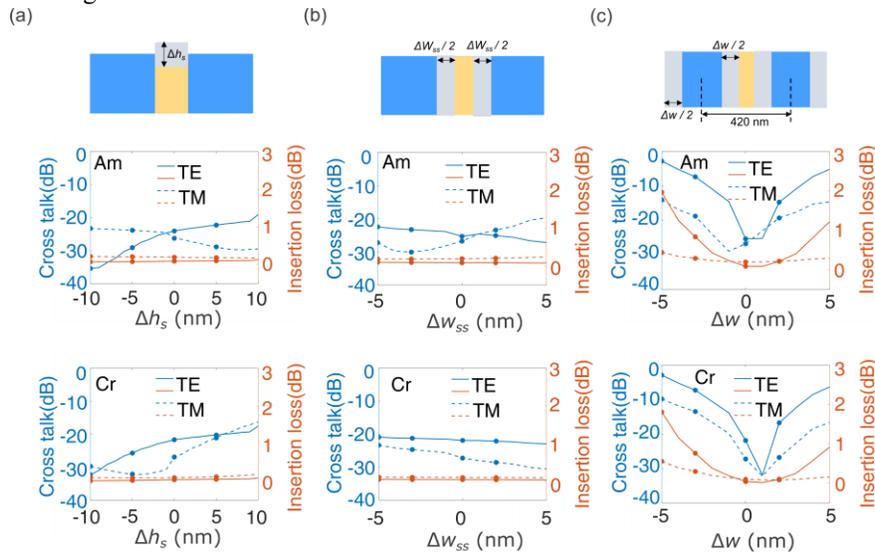

**Figure S4** The influence of fabrication errors on the crosstalk and insertion loss of the polarization-independent switches, including (a) variation in $Sb_2S_3$ height $\Delta h_s$, (b) variation in $Sb_2S_3$ width $\Delta w_{ss}$, (c) variation in waveguide width ($\Delta w$).